# Optimizing food taste sensory evaluation through neural network-based taste electroencephalogram channel selection

Xiuxin Xia [a, b], Qun Wang [a, b], He Wang [a, b], Chenrui Liu [a, b], Pengwei Li [a, b], Yan Shi [a, b], Hong Men [a, b, *]

[a] *School of Automation Engineering, Northeast Electric Power University, Jilin, 132012, China*

[b] *Institute of Advanced Sensor Technology, Northeast Electric Power University, Jilin 132012, China*

\* *Corresponding author.*

Corresponding author.

E-mail address: menhong@neepu.edu.cn (Hong Men).

The e-mail address of each author.

1202200014@neepu.edu.cn (Xiuxin Xia)

2202300723@neepu.edu.cn (Qun Wang)

1202300098@neepu.edu.cn (He Wang)

2202100634@neepu.edu.cn (Chenrui Liu)

1202100044@neepu.edu.cn (Pengwei Li)

shiyan@neepu.edu.cn (Yan Shi)

**Abstract**

The taste electroencephalogram (EEG) evoked by the taste stimulation can reflect different brain patterns and be used in applications such as sensory evaluation of food. However, considering the computational cost and efficiency, EEG data with many channels has to face the critical issue of channel selection. This paper proposed a channel selection method called class activation mapping with attention (CAM-Attention). The CAM-Attention method combined a convolutional neural network with channel and spatial attention (CNN-CSA) model with a gradient-weighted class activation mapping (Grad-CAM) model. The CNN-CSA model exploited key features in EEG data by attention mechanism, and the Grad-CAM model effectively realized the visualization of feature regions. Then, channel selection was effectively implemented based on feature regions. Finally, the CAM-Attention method reduced the computational burden of taste EEG recognition and effectively distinguished the four tastes. In short, it has excellent recognition performance and provides effective technical support for taste sensory evaluation.

## 1. Introduction

Brain activity can be measured using electrocorticography, electroencephalography (EEG), and functional near-infrared spectroscopy (Scholkmann et al., 2014; Thut & Miniussi, 2009; Todaro et al., 2019). The above methods explore the relationship between brain activity, behavior, and psychology by converting neural activity into processable signals. Among them, EEG has been widely used in neuroscience research (Kroupi et al., 2015), disease treatment (Chaudhary et al., 2016), brain-computer interface (Hsu, 2011; Zhang et al., 2018), and other fields (Xia et al., 2023a; Xia et al., 2023b) due to its low cost and high portability.

Food sensory evaluation involves psychology, physiology, and statistics (Vivek et al., 2020), and the evaluation aspects mainly include color, smell, and taste. Among them, food taste is a significant factor. However, most sensory evaluations of food taste currently rely on artificial sensory evaluation and machine perception, which have significant limitations. In artificial sensory evaluation, the results are highly subjective, leading to poor experiment reliability and difficulty reproducing (Espinoza Mina & Gallegos Barzola, 2018). In machine perception, such as electronic tongues, it is challenging to meet the needs of different types of new products due to the limitation of the type and number of sensors (Zheng et al., 2022). Taste EEG can objectively reflect a series of psychological and physiological activities of people after tasting taste, which includes taste information and human sensory perception. It is more objective than artificial sensory evaluation and flexible than machine perception. Therefore, taste EEG has unique advantages for taste sensory evaluation (Xia et al., 2024a). In some related studies, Crouzet et al. decoded the taste EEG of salt, sweet, sour, and bitter through multivariate pattern analysis (Crouzet et al., 2015).

Hashida et al. distinguished the difference between sweet, salty, and water using EEG features extracted by adaptive Gabor transform (Hashida et al., 2005). Domracheva et al. used taste EEG and visual EEG to assess different products' similarities (Domracheva & Kulikova, 2020). Chandran et al. took EEG's mean, median, and power spectral density as features and used support vector machines to classify salt, sweet, sour, and bitter (Chandran & Perumalsamy, 2023). However, most taste EEG recognition uses the traditional machine learning method with manual feature extraction, which reduces recognition efficiency. In addition, the recognition effect largely depended on the extracted features, which made the model less adaptive.

Convolutional neural networks (CNN) have been widely used in EEG decoding, image recognition, and natural speech processing due to their strong spatial feature extraction ability and end-to-end processing flow (Li et al., 2024; Liu et al., 2021; Wang et al., 2024a; Xia et al., 2024b). It has gradually developed from simple models such as early LeNet5 (LeCun et al., 1998) and AlexNet (Krizhevsky et al., 2017) to deeper network layers and more complex network structures to mine the feature information in the data (He et al., 2016; Simonyan & Zisserman, 2014; Szegedy et al., 2016). Deep CNN can extract global and high-level features in deeper layers after extracting local low-level features from EEG data. Therefore, the application of deep CNN in complex EEG tasks is gradually increasing (Lawhern et al., 2018; Schirrmeister et al., 2017). However, the increase in network complexity also inevitably results in information redundancy. Attention mechanisms can better focus on important feature information while effectively ignoring redundant information, so it is widely used in CNN (He et al., 2021; Men et al., 2021; Wang et al., 2020; Woo et al., 2018). In EEG research, Li et al. proposed a multi-scale fusion convolutional neural network based on an attention mechanism to extract multi-scale spatiotemporal features in EEG signals and improve the network's sensitivity (Li et al., 2020). Zhang et al. integrated gender and age factors into a 1D convolutional neural network through an attention mechanism to enable the network to explore complex correlations between EEG signals and demographic factors (Zhang et al., 2020). Jia et al. proposed a spatial-spectral-temporal-based attention 3D dense network for EEG emotion recognition, which can adaptively explore discriminative local patterns by the attention mechanism (Jia et al., 2020).

EEG signals are collected by electrodes placed on the scalp, which usually cover the whole brain. On the one hand, not all electrodes are beneficial in different EEG tasks. When the unfavorable electrode is used in EEG analysis, it will cause information redundancy and increase computational complexity. On the other hand, the decrease in the number of channels will reduce the EEG's spatial resolution, which may lead to a decrease in recognition accuracy. Therefore, adopting an effective channel selection method in EEG tasks is essential. In the related research, Wang et al. extracted the power spectrum and wavelet coefficients of EEG and then selected channels based on the Lasso algorithm (Wang et al., 2015). Arvaneh et al. used a sparse public space pattern algorithm to realize channel selection effectively (Arvaneh et al., 2011). Wang et al. proposed an automatic channel selection method called motor imagery brain-machine interfaces (MI-BMInet), effectively realizing the classification task in ultra-low power MCU based on a spatial filter (Wang et al., 2024b). Wang et al. used normalized mutual information (NMI) to establish the connection matrix between channels and then effectively realized channel selection through thresholding and other operations (Wang et al., 2019). In previous studies, the effectiveness of most channel selection methods was only evaluated by the classification results, and their interpretability was poor.

Gradient-weighted class activation mapping (Grad-CAM) (Selvaraju et al., 2017) is used to visualize the recognition process of CNN. In the image classification task, Grad-CAM can locate the sensitive areas of each class in the image, which effectively visualizes the key areas of CNN feature extraction. EEG was composed of the time and channel dimensions, which is similar to the image in the data structure. So, EEG can be regarded as a special kind of image. Therefore, it is feasible to use Grad-CAM to visualize EEG and then use the visualized EEG to select channels. Finally, the channel selection is realized, and the method is interpretable. In related research, Li et al. used the Grad-CAM method to select channels, which can extract key channel information and realize the balance between the model performance and the number of channels (Li et al., 2020b). In addition, this method combines feature visualization technology and can explain the decision-making process of the deep learning model. However, this method combines recurrent neural networks with CNN for feature mining, and its network structure is relatively complex, requiring more training resources. In addition, it is difficult for the conventional convolutional neural network to pay attention to the key features of EEG, which is not conducive to the visualization of the Grad-CAM method, thus affecting the channel screening effect. In contrast, the method proposed in this work combines the Grad-CAM model and CNN-CSA model. It pays more attention to the key features of EEG by introducing an attention mechanism into CNN, which can make the Grad-CAM method better visualize the important EEG features and thus better realize the selection of EEG channels.

This paper proposed a channel selection method called class activation mapping with attention (CAM-Attention) in taste EEG recognition, which combined the Grad-CAM model and convolutional neural network with the channel and spatial attention (CNN-CSA) model. The contributions are summarized as follows:

(1) The taste EEG experiment was designed, and the taste EEG was collected using different taste stimuli.

(2) To fully exploit the essential features of the taste EEG, the CNN-CSA model was proposed.

(3) The CNN-CSA and the Grad-CAM models were combined to form the CAM-Attention method to effectively select the key channels in the taste EEG, then the CNN-CSA model was used to mine the data features in the selected channels. Finally, the difference between the four tastes was effectively distinguished while reducing the amount of calculation. The above method can provide technical support for the related research on taste sensory evaluation.

**2. Materials and methods**

**2.1 Experimental and data**

**2.1.1 Subjects**

The research conformed to the revised Helsinki Declaration, and the program was approved by the Northeast Electric Power University Scientific Research Ethics and Science and Technology Safety and Committee. Before the experiment, 20 right-handed subjects (10 males and 10 females) aged between 20 and 30 were recruited by posting announcements in the school. All subjects had no neurological disease or dysgeusia. They all signed the informed consent form. On the day of the experiment, they were told to wash their hair and brush their teeth (odorless toothpaste) in the morning and not to eat (except water) for two hours before the experiment.

**2.1.2 Materials and instrument**

The four common food tastes, sour, sweet, bitter, and salty, were used as experimental materials to induce taste EEG. According to reference (Wallroth et al., 2018), the four taste solutions were prepared as follows: sour (0.075 g food grade citric acid dissolved in 100 ml distilled water, 0.039 M), sweet (15 g sucrose dissolved in 100 ml distilled water, 0.44 M), bitter (3 g bitter melon powder dissolved in 100 ml distilled water) and salty (3.8 g sodium chloride dissolved in 100 ml distilled water, 0.65 M), in addition, a 100 ml bottle of distilled water was prepared. In the pre-experiment, the prepared solution can be perceived by the subjects without causing them discomfort.

The taste solution in the sample bottle was delivered to the center of the subject's tongue through a self-developed taste inducer with high precision, high stability, and low noise. Its detailed structure is shown in Fig. 1. The taste inducer was mainly composed of an STM32F103ZE control panel (Dongguan Wildfire Electronic Technology Co., Ltd., China), S15S-53J micro vacuum pump (Chengdu Hailin Technology Co., Ltd., China), and SFO-1037V-01 solenoid valve (Dongguan Sifang Electronic Technology Co., Ltd., China). Among them, the control panel could precisely control the delivery rate and time of the taste solution by controlling the micro vacuum pump and solenoid valve. At the same time, the solenoid valve and vacuum pump were wrapped with sound insulation cotton to eliminate noise.

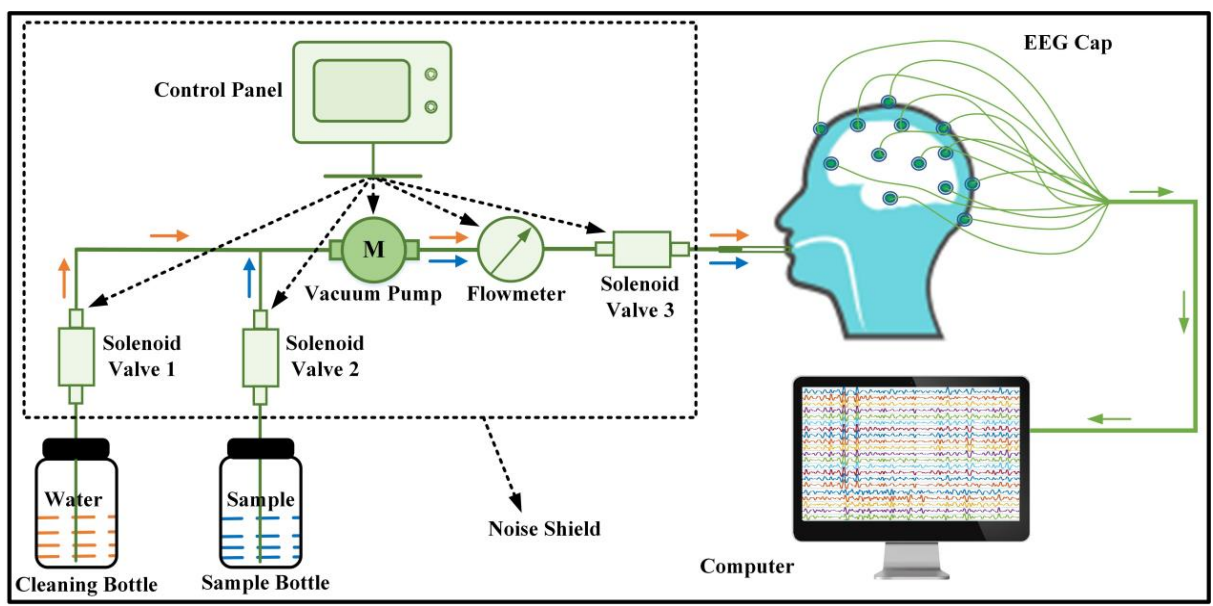

**Fig. 1.** Taste EEG evoked device.

The NCEP-P EEG acquisition system (Shanghai NCC Electronics Co., Ltd., China) was used to acquire taste EEG with a sampling frequency of 256 Hz. According to the 10-20 system, 21 electrodes (Fz、Cz、Pz、T3、T4、C3、C4、Fp1、Fp2、F7、F8、T5、T6、O1、O2、F3、F4、P3、P4、A1、A2) of the EEG cap (Greentek Pty. Ltd., China) were placed in the corresponding positions.

**2.1.3 Taste EEG acquisition process**

Before the experiment, the experimental samples were placed in 250 ml sampling bottles and heated to 37.5 °C. The laboratory temperature was controlled at 21 ± 2 °C and ensured no strong electromagnetic interference was nearby. Subjects were put on EEG caps, earplugs, and nasal plugs. The chin pad was adjusted so that the outlet tube of the taste inducer was located 0.25 - 0.5 cm above the center of the subject's tongue. For each subject, the taste EEG was collected over four days, which were parallel experiments, and each day was from 9:00 am to 11:00 am or 3:00 pm to 5:00 pm. The taste EEG experimental process for each subject on each day is shown in Fig. 2. The order of the four taste experiments was randomized, and the interval between adjacent experiments was 30 minutes to allow the subjects to rinse their mouths and rest adequately. In each taste experiment, the subjects were stimulated by water twice to adapt to the feeling of water flowing on the tongue. For water stimulation, water flows into the center of the subjects' tongues at a constant flow rate of 0.25 ml/s for 2 s. Then, the subjects spat out the liquid within 10 seconds and took a short rest. After two water stimulations, the taste solution flowed into the center of the subjects' tongues at a constant flow rate of 0.25 ml/s for 2 s. Then, the subjects tasted the taste for 10 seconds to induce a taste EEG, during which they closed their eyes lightly without swallowing.

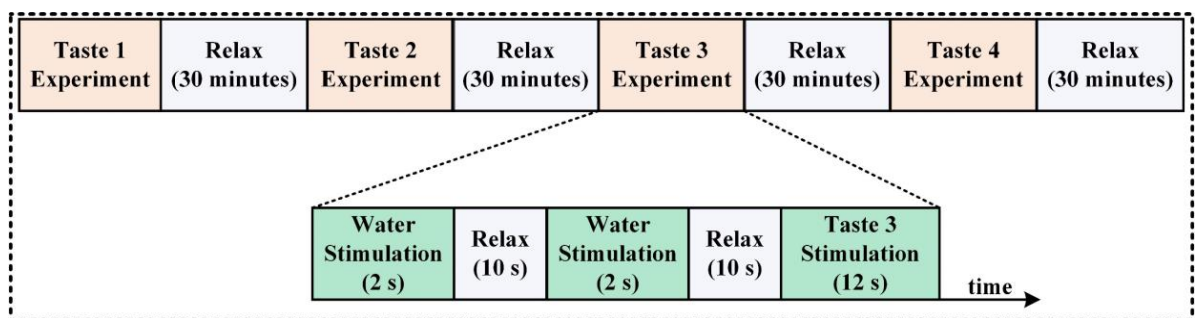

**Fig. 2.** The taste EEG experimental process.

**2.1.4 Preprocessing**

The taste EEG was preprocessed by Matlab (R2017b) and its built-in toolkit EEGLAB (version 2021). The processing flow is as follows.

(1) 320 (20 × 4 × 4) taste EEG data segments were collected, each 10 s size. The sample size of the

taste EEG was set to 2 s, so 1600 taste EEG samples could be generated. The labels of the four taste EEG samples of sour, sweet, bitter, and salty were set as *label0*, *label1*, *label2*, and *label3*, respectively.

(2) After establishing the taste EEG samples, a finite impulse response filter was used to perform bandpass filtering between 0.5 and 50 Hz to remove low-frequency and high-frequency noise in the samples. Notch filters with the lower edge of the 49 Hz passband and the upper edge of the 51 Hz passband were used to remove power frequency noise from the samples. Then, the sampling frequency of the samples was reduced from 256 Hz to 128 Hz to reduce the sample size, so the size of the final EEG sample was 256 × 21.

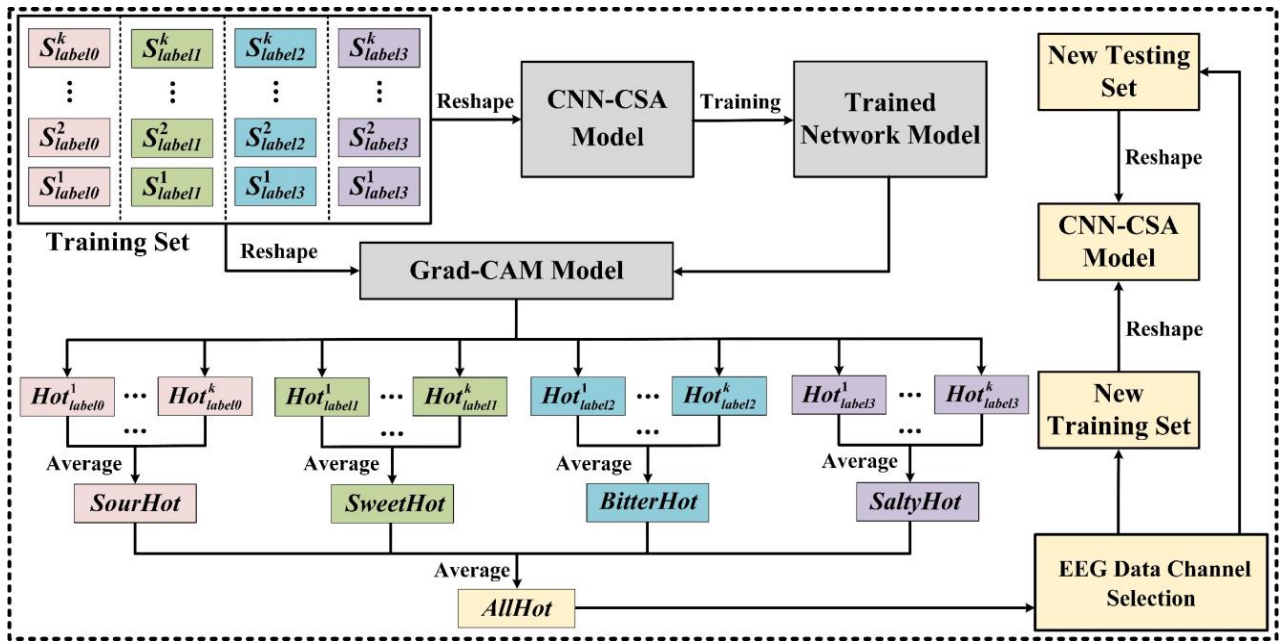

**Fig. 3.** Analysis framework of the CAM-Attention channel selection method.

**2.2 Analytical method**

**2.2.1 Analysis framework**

The analysis framework of the CAM-Attention channel selection method is shown in Fig. 3, and the steps are as follows.

Step 1: Each taste EEG sample in the training set was reshaped into 84 × 64 and then input to the CNN-CSA model for the classification training of sour, sweet, bitter, and salty.

Step 2: The trained CNN-CSA model and the reshaped training samples were input into the Grad-CAM model to generate the gradient class activation map corresponding to each training sample.

Step 3: To obtain the whole gradient class activation map of each class of EEG samples. All the gradient class activation maps in each class were averaged. For example, the sour's average gradient class activation map is calculated as follows.

$$SourHot = \frac{1}{k}\sum_{i}^{k} Hot_{label0}^{i} \tag{1}$$

Where $Hot_{label0}^{i}$ represented the $i$th gradient class activation map of sour, and $k$ was the number of gradient class activation maps in sour.

Step 4: The four tastes' average gradient class activation maps were averaged to generate gradient class activation maps *AllHot* representing the activation regions of the four tastes for EEG channel selection.

Step 5: The sum of EEG each channel's values in *AllHot* was calculated and sorted in descending order, the formula is as follows.

$$ChannelList = \text{sort}([\sum_{n=1}^{256} AllHot_{m}^{n}, \cdots]) \tag{2}$$

Where $m$ in *AllHot* represented the EEG channel numbers 1, 2, ..., 21, respectively. Sort ( ) meant to sort the sum of each EEG channel's values from largest to smallest.

Step 6: Determined the EEG channel number $Q$ and selected the first $Q$ EEG channels in *ChannelList*. In the training and testing sets, only the data in the selected EEG channels were saved and used to generate a new training and testing set. Then, each sample in the new training and testing sets was reshaped into 84 × 64. Finally, the reshaped training set samples were input into the CNN-CSA model for training, and the reshaped testing set samples were input into the trained CNN-CSA model for testing.

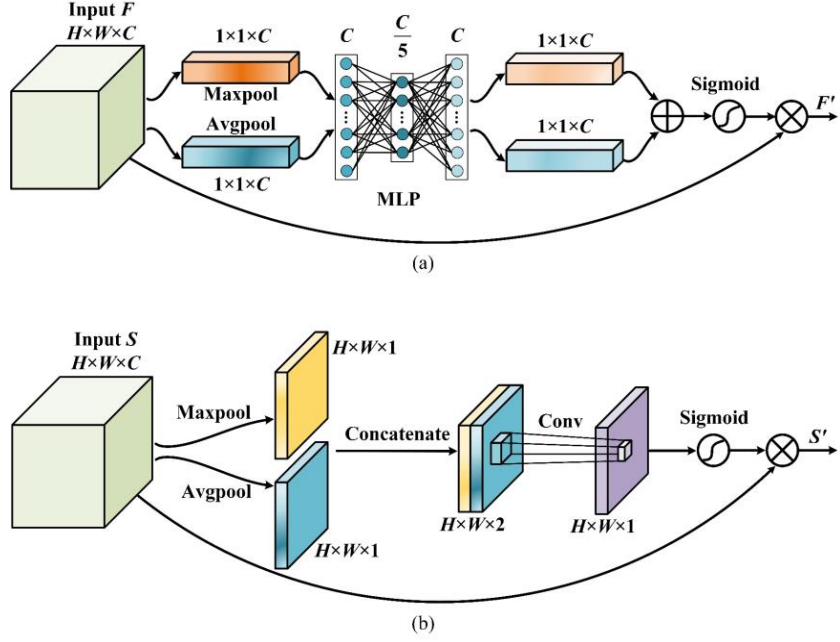

**Fig. 4.** The structures of the attention module: (a) CAM, (b) SAM.

**2.2.2 Channel attention and spatial attention modules**

In this paper, an attention mechanism was introduced into the convolutional neural network, which can effectively extract the essential features of the taste EEG and significantly improve the recognition performance of the network. Channel attention and spatial attention were two important attention mechanisms. Among them, channel attention can pay attention to important channels in the feature map and ignore redundant channels more effectively when extracting deep EEG features. Spatial attention can pay attention to the important spatial position in the feature map and mine the important spatial features more effectively when extracting the shallow EEG features.

The structures of the channel attention module (CAM) and the spatial attention module (SAM) are shown in Fig. 4(a) and (b), respectively. An EEG feature map $F$ with the size $H \times W \times C$ was used as input for CAM. Firstly, the average pooling and max pooling were used to aggregate the spatial information of the input EEG feature map $F$. The aggregated EEG spatial information was respectively input into the shared multi-layer perception with $\frac{C}{5}$ neurons in the hidden layer, and two interactive EEG feature vectors were obtained, with a size of 1 × 1 × C. Then, the EEG feature vectors were added and activated by the sigmoid function to get the weight coefficient representing the importance of each channel in the EEG feature map. Finally, the weight coefficient was multiplied with the input EEG feature map $F$ to obtain a new EEG feature map $F'$ with channel attention.

SAM can express the importance of the EEG feature map in the spatial dimension. An EEG feature map $S$ with the size $H \times W \times C$ was used as input for CAM. Firstly, average pooling and maximum pooling were used to aggregate channel information of EEG in the channel dimension. And the two aggregated feature vectors were stitched together along the channel dimension to form an EEG feature vector with the size of $H \times W \times 2$ to achieve feature fusion. Then the convolution operation was performed on the fused features to achieve feature extraction of the aggregated EEG feature map. The convolution kernel size was 3 × 3, the padding was 1, and the step size was 1. Finally, the convolved feature vector was activated by the sigmoid function and multiplied by the input EEG feature map $S$ to obtain a new EEG feature map $S'$ with spatial attention.

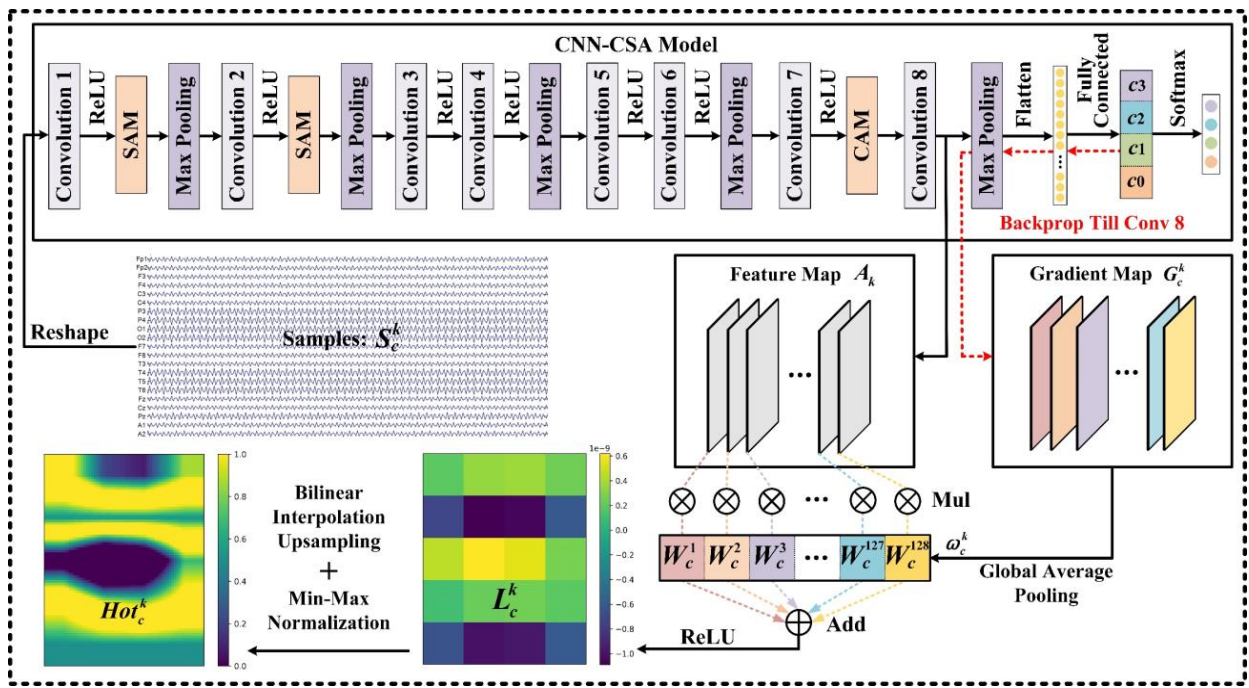

**Fig. 5.** The generation process of the gradient class activation map.

Table 1 The parameters of the CNN-CSA model.

| No. | Operating | Kernel size | Stride | Padding | Input size | Output size | Active function |
|-----|-----------|-------------|--------|---------|------------|-------------|-----------------|
| 1 | Convolution 1 | 3×3 | 1 | 1 | 84×64×1 | 84×64×8 | ReLU |
| 2 | SAM | — | — | — | 84×64×8 | 84×64×8 | — |
| 3 | Max pooling | 2×2 | 2 | 0 | 84×64×8 | 42×32×8 | — |
| 4 | Convolution 2 | 3×3 | 1 | 1 | 42×32×8 | 42×32×16 | ReLU |
| 5 | SAM | — | — | — | 42×32×16 | 42×32×16 | — |
| 6 | Max pooling | 2×2 | 2 | 0 | 42×32×16 | 21×16×16 | — |
| 7 | Convolution 3 | 3×3 | 1 | 1 | 21×16×16 | 21×16×32 | ReLU |
| 8 | Convolution 4 | 3×3 | 1 | 1 | 21×16×32 | 21×16×32 | ReLU |
| 9 | Max pooling | 2×2 | 2 | 0 | 21×16×32 | 10×8×32 | — |
| 10 | Convolution 5 | 3×3 | 1 | 1 | 10×8×32 | 10×8×64 | ReLU |
| 11 | Convolution 6 | 3×3 | 1 | 1 | 10×8×64 | 10×8×64 | ReLU |
| 12 | Max pooling | 2×2 | 2 | 0 | 10×8×64 | 5×4×64 | — |
| 13 | Convolution 7 | 3×3 | 1 | 1 | 5×4×64 | 5×4×128 | ReLU |
| 14 | CAM | — | — | — | 5×4×128 | 5×4×128 | — |
| 15 | Convolution 8 | 3×3 | 1 | 1 | 5×4×128 | 5×4×128 | — |
| 16 | Max pooling | 2×2 | 2 | 0 | 2×2×128 | 2×2×128 | — |
| 17 | Flatten | — | — | — | 2×2×128 | 512 | — |
| 18 | Fully connected | — | — | — | 512 | 4 | — |
| 19 | Softmax | — | — | — | 4 | 4 | — |

**2.2.3 CNN-CSA model**

CNN can effectively extract high-dimensional features of EEG data by convolution and pooling. Therefore, in recent years, CNN has been widely used in EEG classification tasks and has shown significant advantages. The CNN-CSA model was proposed in this paper to effectively implement channel selection and taste recognition in taste EEG tasks. Its structure is shown in Fig. 5. The CNN-CSA model consisted of convolution, pooling, a fully connected layer, a softmax layer, and attention modules. First, the shallow features of the taste EEG data were extracted by convolution 1. Compared with the deep EEG feature map after feature extraction, the shallow EEG feature map contained more spatial redundant information, so the SAM focused on the essential spatial regions in the shallow EEG feature map. Similarly, after convolution 2, the SAM was used to further extract the information of important spatial regions in the shallow EEG feature map. Then, the deep EEG features were gradually extracted by a series of convolution and pooling. After convolution 7, the number of channels of the EEG feature map was expanded to 128. With the deepening of the network depth, there were inevitably redundant channels in the deep EEG feature maps. So CAM was used to effectively focus on the important channels and ignore the redundant channels in the deep EEG feature maps. Finally, the feature map with channel attention was flattened after convolution 8 and pooling, and then the final prediction output was obtained after the fully connected layer and softmax layer. The detailed parameters of the CNN-CSA model are shown in Table 1.

**2.2.4 Grad-CAM model**

Many studies have shown that CNN can extract deep features through convolution and pooling, expressing deeper data information. In addition, convolution and pooling can keep the spatial information of data well. Therefore, compared with the front-end convolution layer and the back-end full connection layer, the last convolution layer in CNN can not only express the deep EEG feature information but also retain the spatial information of the EEG data. Grad-CAM visually explains the classification and recognition of CNN using gradient information back-propagating from the class score to the last convolution layer. In this work, Grad-CAM and CNN-CSA were used to generate gradient class activation maps for different classes of taste EEG. Its process is shown in Fig. 5.

(1) The convolution process means the convolution kernel's movement. Suppose the original EEG sample size of 21 × 256 is directly input into the CNN-CSA model for training. The model will mine more EEG features between channels during the convolution process, which is not conducive to generating the gradient class activation map for observing the contribution of a single EEG channel to the classification. To explore more features between the EEG data in each channel, the original EEG sample size of 21 × 256 was reshaped into 84 × 64 before being input into the CNN-CSA model for training.

(2) For example, after the taste EEG sample $S_c^k$ was reshaped into 84 × 64, it was propagated forward to the softmax layer in the trained CNN-CSA model, and the scores of each class were obtained. It was worth noting that the channel and spatial attention modules were introduced into the CNN-CSA model to better mine the feature information of the EEG samples, ensuring the effectiveness of the gradient class activation map. Then calculating the gradient of the score $y_c$ of class $c$ before the softmax layer relative to the EEG feature map $A_k$ after the last convolution layer. The calculation formula is as follows.

$$G_c^k = \frac{\partial y^c}{\partial A_k} \tag{3}$$

Where $G_c^k$ represents the gradient map of class $c$ score on the EEG feature map $A_k$.

(3) Global average pooling was performed on the gradient map along the channel dimension to obtain the weight $\omega_c^k$ representing the overall gradient value of each channel in the gradient map. The calculation formula is as follows.

$$\omega_c^k = \frac{1}{H \times W} \sum_{i=1}^{H} \sum_{j=1}^{W} G_{ij} \tag{4}$$

where $G_{ij}$ represents the feature of each channel in $G_c^k$, and $H$ and $W$ represent the height and width of the $G_{ij}$.

(4) The weight was multiplied by the corresponding channel in EEG feature map $A_k$ and added according to the corresponding positions along the channel dimension. Then, the ReLU activation function was used to eliminate the interference of irrelevant classes, highlighting class $c$, and the low-resolution gradient class activation map $L_c^k$ with a size of 5 × 4 was obtained. Its calculation formula is as follows.

$$L_c^k = \mathrm{ReLU}(\sum_k \omega_c^k \times A_k) \tag{5}$$

(5) To observe the heat values of different EEG channels in the gradient class activation map, upsampling the low-resolution gradient class activation map $L_c^k$ to the size of EEG samples by bilinear interpolation. After upsampling, it was normalized by maximum and minimum to unify the numerical range in the gradient class activation map of taste EEG. Finally, the gradient class activation map $Hot_c^k$ of size 84 × 64 was obtained by min-max normalization.

## 3. Results and discussion

## 3.1 Model setting and evaluation method

All taste EEG samples were randomly divided into the training set and testing set according to 3 : 1, so the number of samples in the training set and testing set were 1200 and 400, respectively. After parameter pre-adjustment, the batch sizes of the training set and testing set of the CNN-CSA model were 64 and 32, respectively. Adam optimizer was used. The learning rate was 0.001, the weight decay was 0.005, and the number of iterations was 200. Accuracy and F1-score were used as evaluation indexes to evaluate the effectiveness of the channel selection method and the CNN-CSA model's classification performance.

## 3.2 Performance evaluation of channel selection method

In this section, the effectiveness of CAM-Attention and other classical channel selection methods such as Lasso (Wang et al., 2015), MI-BMInet (Wang et al., 2024b), and NMI (Wang et al., 2019) were evaluated. CAM-Attention is combined with the CNN-CSA model for taste EEG recognition. To ensure the reliability of the experiments, five independent experiments were performed for each channel selection method, and their accuracy and F1-score were averaged.

Table 2 shows the classification results of different channel selection methods. The CAM-Attention method outperformed other channel selection methods in recognizing taste EEG. Furthermore, Lasso performed the worst, and NMI and MI-BMInet performed comparably. For the above methods, with the reduction of the number of selected channels, the classification effects of the four methods deteriorated to varying degrees. When the number of selected channels was 12, the accuracy of Lasso, NMI, and MI-BMInet was 80.26%, 87.49%, and 88.38%, respectively, which was 4.10%, 1.93%, and 3.68% lower than that of all channels. In this case, although Lasso, NMI, and MI-BMInet effectively reduce the amount of computation, their classification performance was significantly worse. For the CAM-Attention method, when the number of selected channels was 12, the accuracy and F1-score were 97.55% and 97.52%, respectively, 0.44% and 0.43% lower than all channels, respectively. It can be seen that the CAM-Attention method effectively realized the selection of essential channels and ensured an excellent classification effect while ensuring a large reduction in the amount of calculation.

Table 2 Classification results of different channel selection methods.

| Channel | Lasso | | NMI | | MI-BMInet | | CAM-Attention | |
| --- | --- | --- | --- | --- | --- | --- | --- | --- |
| | Accuracy | F1-score | Accuracy | F1-score | Accuracy | F1-score | Accuracy | F1-score |
| 4 | 59.41 | 58.52 | 62.72 | 62.19 | 53.70 | 52.24 | 80.52 | 80.04 |
| 8 | 75.07 | 74.56 | 73.62 | 73.80 | 80.32 | 79.20 | 93.47 | 93.18 |
| 12 | 80.26 | 80.07 | 87.49 | 87.96 | 88.38 | 87.83 | 97.55 | 97.52 |
| 16 | 81.24 | 81.01 | 88.72 | 89.66 | 88.38 | 87.91 | 97.80 | 97.79 |
| 21 | 84.36 | 84.25 | 89.42 | 90.34 | 92.06 | 91.69 | 97.99 | 97.95 |

The line chart reflecting the performance of different channel selection methods is shown in Fig. 6. From Fig. 6(a) and (b), it can be seen that the CAM-Attention method outperformed other models in both accuracy and F1-score when selecting all or part of the channels. When the number of selected channels was less than 12, the recognition accuracy and F1-score of the above four methods significantly dropped, indicating that too much reduction in the number of channels would lead to the loss of important information in taste EEG. The loss of important information was not conducive to recognizing taste EEG. Therefore, choosing 12 channels was ideal compared to selecting 4, 8, 16, and 21 channels.

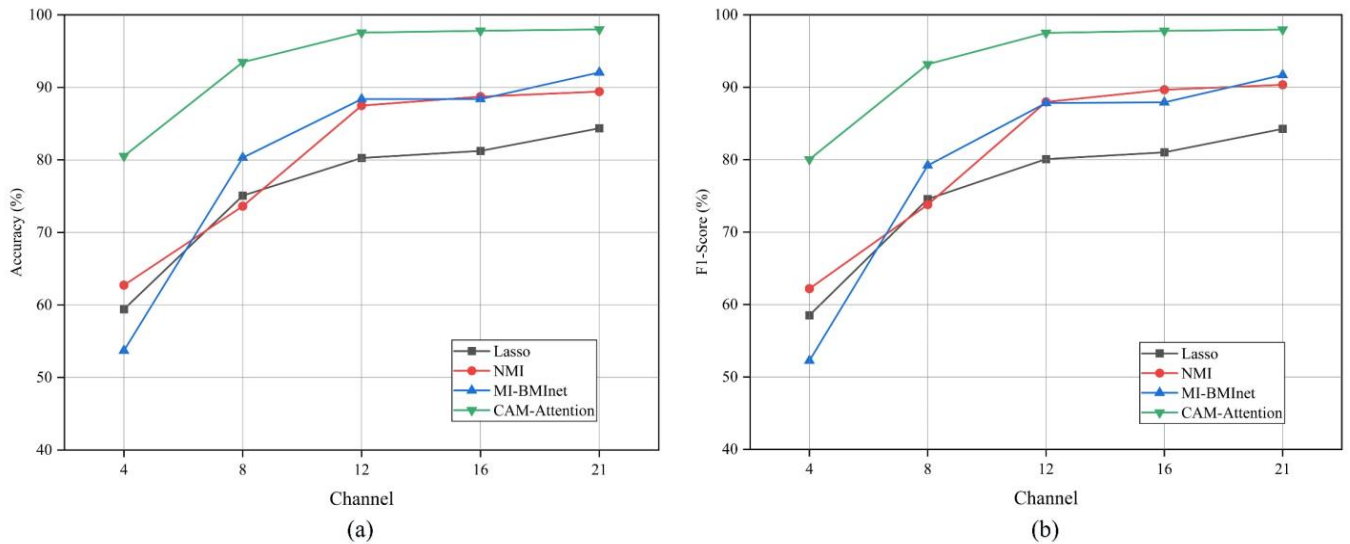

**Fig. 6.** Effectiveness evaluation of different channel selection methods: (a) accuracy, (b) F1-score.

### 3.3 Evaluating different network models combined with Grad-CAM

The Grad-CAM and CNN-CSA models were two important parts of the CAM-Attention methods. Among them, the change in the network structure of the CNN-CSA model would lead to a change in the gradient class activation map. The shallow network can save computing resources, while the deep network has more advantages in deep feature extraction. In this section, we selected the shallow classical networks LeNet5 (LeCun et al., 1998), AlexNet (Krizhevsky et al., 2017), EEGNet (Lawhern et al., 2018), ViT (Dosovitskiy et al., 2020), and deeper ResNet18 (He et al., 2016) networks, which have been widely used in EEG analysis to explore the effect of the networks combined with the Grad-CAM model. Then, the effectiveness of the CNN-CSA model combined with the Grad-CAM model was fully proved by comparing the effects of the above classical networks with those of the CNN-CSA model. Among them, only the CNN-CSA model in the CAM-Attention method was replaced with the different network model. The batch size, optimizer, learning rate, weight decay, and iteration times of all network models were the same as those in the CAM-Attention method. To ensure the reliability of the experiments, five independent experiments were carried out on the channel selection methods under each network model, and their accuracy and F1-score were averaged.

Table 3 Classification results of Grad-CAM combined with different networks for channel selection.

| Channel | EEGNet | | ResNet18 | | LeNet5 | | ViT | | AlexNet | | CNN-CSA | |
|---|---|---|---|---|---|---|---|---|---|---|---|---|
| | Accuracy | F1-score | Accuracy | F1-score | Accuracy | F1-score | Accuracy | F1-score | Accuracy | F1-score | Accuracy | F1-score |
| 4 | 33.40 | 30.98 | 49.97 | 47.11 | 60.32 | 59.03 | 48.57 | 47.27 | 73.67 | 72.23 | 80.52 | 80.04 |
| 8 | 49.27 | 46.92 | 59.87 | 57.70 | 76.83 | 75.96 | 73.65 | 72.82 | 92.09 | 92.23 | 93.47 | 93.18 |
| 12 | 57.27 | 55.97 | 69.59 | 68.10 | 84.76 | 84.23 | 87.05 | 86.01 | 94.99 | 94.84 | 97.55 | 97.52 |
| 16 | 61.59 | 59.65 | 76.57 | 75.70 | 86.03 | 85.26 | 94.92 | 94.72 | 95.78 | 95.51 | 97.81 | 97.79 |
| 21 | 69.78 | 68.02 | 88.25 | 87.99 | 90.03 | 89.62 | 95.75 | 95.4 | 96.94 | 96.88 | 97.99 | 97.95 |

Table 3 shows the results of different networks combined with Grad-CAM for channel selection. Compared to other network models, the CNN-CSA model achieved the best results when combined with the Grad-CAM model. Without employing any EEG channel selection, the CNN-CSA model achieved an accuracy of 97.99% and an F1-score of 97.95%. These figures surpassed those of other models, including AlexNet by 1.05% and 1.07%, ViT by 2.24% and 2.55%, LeNet5 by 7.96% and 8.33%, ResNet18 by 9.74% and 9.96%, and EEGNet by 28.21% and 29.93%, respectively. It shows that the CNN-CSA model has advantages in extracting key features from taste EEG data after combining the above network model with the Grad-CAM model to select channels. When the number of EEG channels was reduced from 21 to 12, a noticeable decline in accuracy and F1-score was observed across various models: EEGNet, ResNet18, LeNet5, ViT, and AlexNet. Specifically, the accuracy decreased by 12.51%, 18.66%, 5.27%, 8.70%, and 1.95%, respectively, while the F1-score dropped by 12.05%, 19.89%, 5.39%, 9.39%, and 2.04%, respectively. However, the CNN-CSA model has hardly decreased in accuracy and F1-score. It underscores the importance of combining the CNN-CSA and Grad-CAM models in preserving critical EEG channels during EEG channel selection.

In addition, we found that the CNN-CSA model achieved better results than the ViT model based on transformer architecture. We think there are the following reasons: (1) Local feature extraction ability: The CNN-CSA model is more suitable for extracting features from local areas in the design. In taste EEG data, some local electrical signal patterns may be associated with specific tastes. The CNN-CSA model can effectively capture these local features through convolution operation to better distinguish different taste stimuli. (2) Data efficiency: Because the taste EEG data is usually high-dimensional and complex, the ViT model may need more training data to train the model effectively. However, the CNN-CSA model can use limited data better to learn effective feature representation because it is more sensitive to capturing local features. (3) Combining the advantages of Grad-CAM: CNN-CSA and Grad-CAM models can better select the key EEG channels. Grad-CAM helps to determine the electrode channels with important information in taste EEG data by visualizing the attention region of the model. In contrast, the CNN-CSA model can extract and classify features on these key channels more effectively. To sum up, the advantages of the CNN-CSA model compared with the ViT model in the task of taste EEG data are mainly reflected in the better capture of local features, higher data efficiency, and the advantages when combined with the Grad-CAM model. These characteristics make the CNN-CSA model perform better in electrode selection and identification of taste EEG data.

Table 4 shows the params and flops of different network models with different numbers of channels. Compared with the ResNet18 and CNN-CSA models, the number of neurons in the EEGNet, LeNet5, and AlexNet models in the first fully connected layer increased with the sample size, so the network parameters also increased with the increase of channel number. With the decrease in the number of channels, the flops of the above models were obviously reduced, which indicated that reducing the number of channels can effectively improve the recognition efficiency of taste EEG. For the CNN-CSA model, reducing the number of EEG channels from 21 to 12 led to a 41.18% decrease in flops. Despite this reduction, the recognition accuracy and F1-score of taste EEG only decreased by 0.44% and 0.43%, respectively. These findings demonstrate the effectiveness of combining the CNN-CSA model with the Grad-CAM model in performing EEG channel selection, thereby enhancing the recognition efficiency of taste EEG signals.

Table 4 Params and flops under different networks and different channel numbers.

| Channel | EEGNet | | ResNet18 | | LeNet5 | | ViT | | AlexNet | | CNN-CSA | |
|---|---|---|---|---|---|---|---|---|---|---|---|---|
| | Params | Flops | Params | Flops | Params | Flops | Params | Flops | Params | Flops | Params | Flops |
| 4 | 0.29M | 1.37M | 11.19M | 33.97M | 1.86M | 4.10M | 3.10M | 3.20G | 0.95M | 1.87M | 1.20M | 3.07M |
| 8 | 0.36M | 3.46M | 11.19M | 67.94M | 3.34M | 8.16M | 3.10M | 6.31G | 1.34M | 4.75M | 1.20M | 6.13M |
| 12 | 0.46M | 6.14M | 11.19M | 101.90M | 4.81M | 12.22M | 3.10M | 9.44G | 1.74M | 7.63M | 1.20M | 9.20M |
| 16 | 0.61M | 10.28M | 11.19M | 135.87M | 6.29M | 16.28M | 3.10M | 12.57G | 2.13M | 10.50M | 1.20M | 12.26M |
| 21 | 0.68M | 14.08M | 11.19M | 186.74M | 7.76M | 21.26M | 3.10M | 16.49G | 2.72M | 14.48M | 1.20M | 15.64M |

Fig. 7 illustrates a line graph depicting the performance of various networks when combined with Grad-CAM. The graph demonstrates that the combination of Grad-CAM with the CNN-CSA model consistently outperformed other configurations in terms of accuracy and F1-score, regardless of whether all or part of the channels were selected. Furthermore, as the number of channels decreased, the CNN-CSA model exhibited a smaller decrease in accuracy and F1-score. It suggests that the CNN-CSA model is more adept at extracting crucial information from key channels and mitigating information loss during the channel selection.

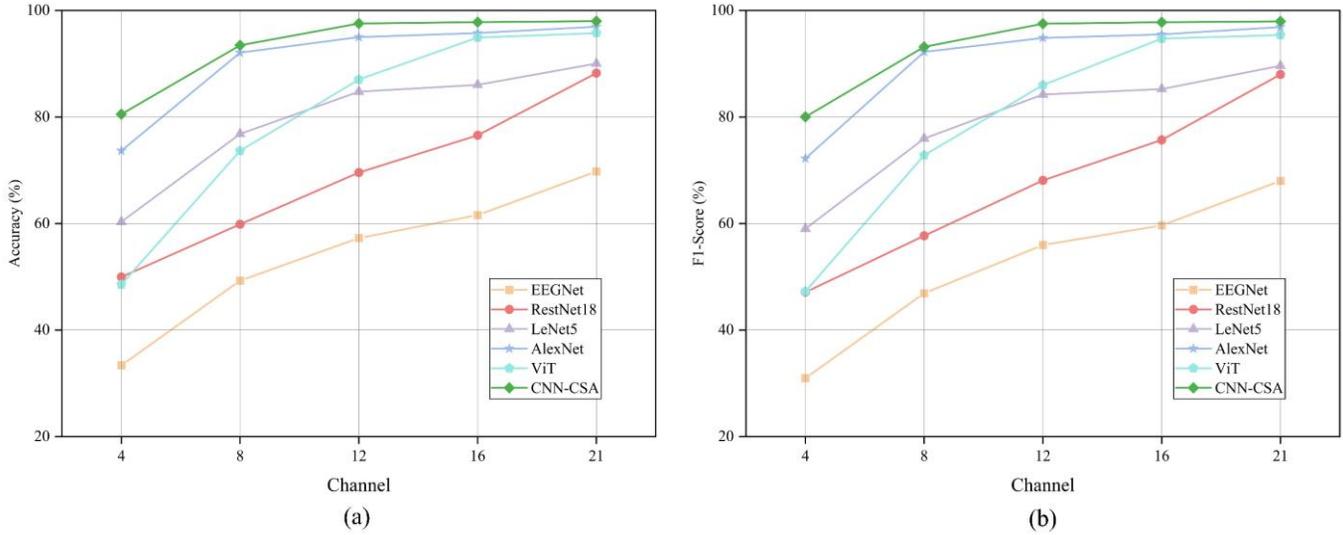

**Fig. 7.** Performance evaluation of different networks combined with Grad-CAM: (a) accuracy, (b) F1-score.

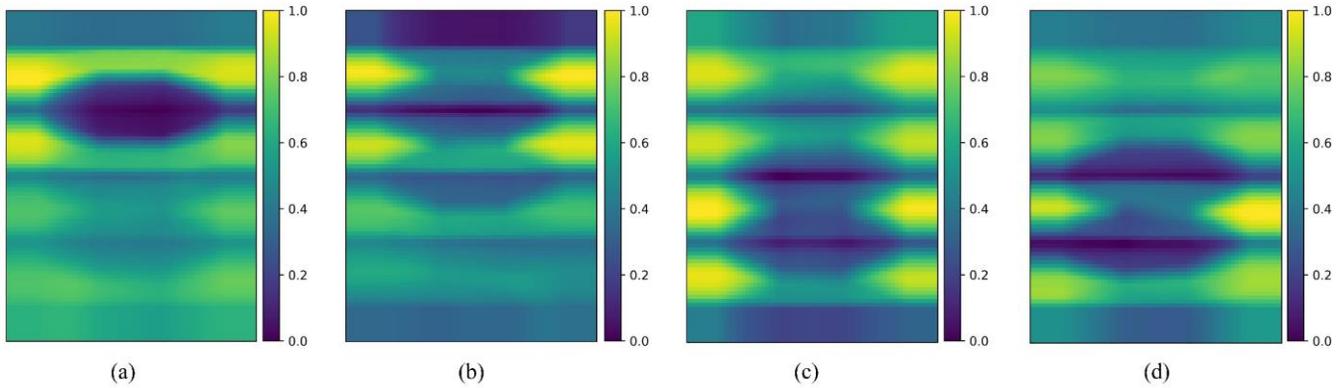

**Fig. 8.** Average gradient class activation map: (a) sour, (b) sweet, (c) bitter, and (d) salty.

**3.4 Brain region distribution of activation maps for different tastes**

The average gradient class activation map of each taste calculated according to Fig. 3 is shown in Fig. 8. The size of each average gradient class activation map was 84 × 64. For the 84 rows of data of the average gradient class activation map, every four rows from top to bottom were the average gradient class activation areas corresponding to one channel. The horizontal dimension of the average gradient activation map represented the time dimension, and the vertical dimension represented the space dimension (channel dimension). It can be seen that the average gradient class activation maps of different tastes were different locally, but the main activation areas were roughly similar. At the same time, the average gradient class activation maps of bitter and salty taste were similar. Still, the average gradient class activation maps of sweet and salty tastes were quite different, which may be used to reflect the differences in subjects' feelings caused by different tastes in future research.

The values of each channel in each average gradient class activation map were summed up separately, and the first 12 channels with the largest sum in each map were taken. Then the electrodes corresponding to the channels were marked in the 10-20 system brain electrode distribution diagram. Finally, the feature channels activated by different tastes are shown in Fig. 9. Overall, the feature channels were mainly concentrated in the middle and back of the brain. The common feature channels activated by the four tastes were distributed in the middle of the frontal lobe, the front of the temporal lobe, the occipital lobe, and the central part of the brain. At the same time, the distribution of the feature channels activated by the four tastes was quite different in the parietal lobe and the posterior part of the temporal lobe, which reflected that the brain regions activated by different tastes were different.

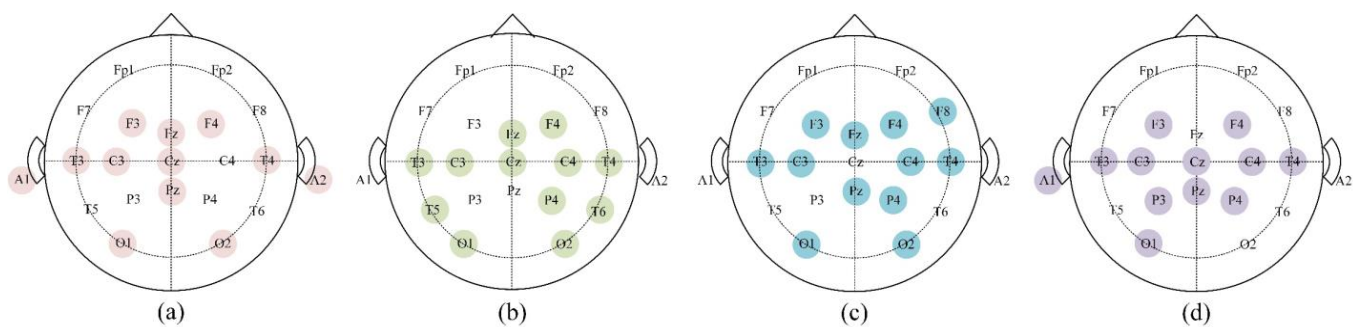

**Fig. 9.** Feature channels activated under different tastes: (a) sour, (b) sweet, (c) bitter, and (d) salty.

**3.5 Cross-time taste EEG recognition**

Due to the physiological and psychological differences in individuals across different periods, recognition models' stability and generalization capability face significant challenges in cross-time taste EEG recognition. Additionally, taste EEG signals are weak and susceptible to noise interference, making cross-time taste EEG recognition extremely challenging. Nevertheless, to further validate the effectiveness of the proposed CAM-Attention method and the CNN-CSA model, we conducted cross-time taste EEG recognition. In this study, we utilized all participants' data from the first three days as the training set and the last day as the test set, resulting in 1200 and 400 samples in the training and test sets, respectively.

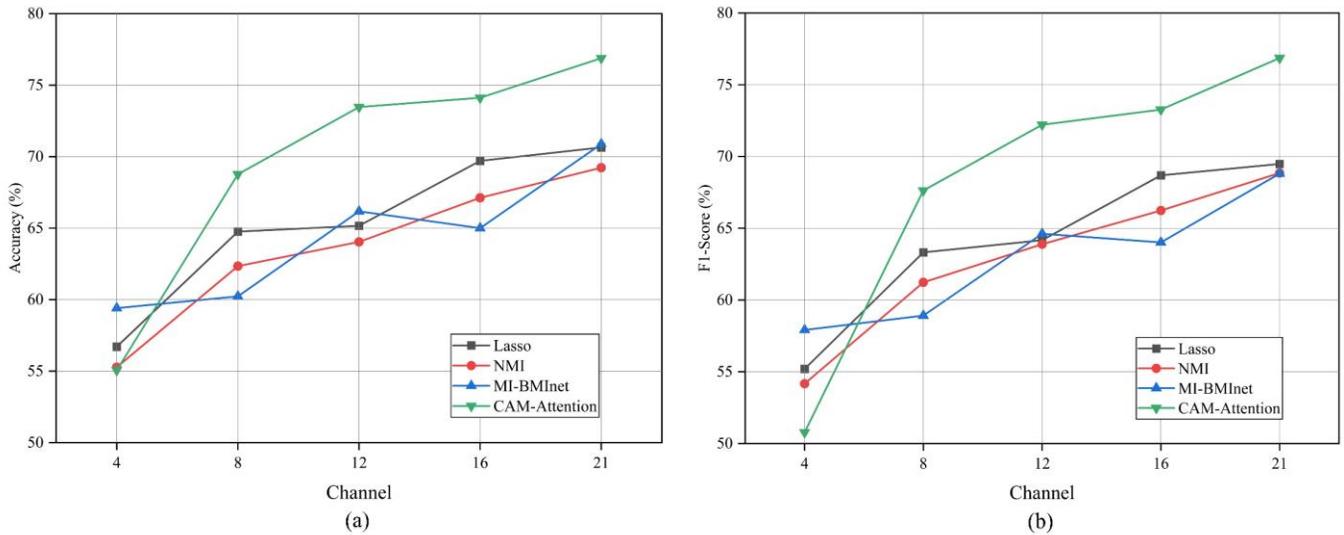

**Fig. 10.** Evaluation of EEG channel selection methods in cross-time EEG recognition: (a) accuracy, (b) F1-score.

Fig. 10 shows the performance comparison of various channel selection methods in cross-time taste EEG recognition, and the configuration of the methods is consistent with that in section 3.2. Notably, when reducing the number of channels from 21 to 12, only the CAM-Attention method sustains robust recognition performance, indicating it effectively selects the important channels representing the taste information in the cross-time taste EEG recognition. Furthermore, comparative analysis reveals that the CAM-Attention method consistently outperforms alternative approaches in cross-time taste EEG recognition, which shows the advantages of the proposed CAM-Attention method in cross-time taste EEG recognition.

Fig. 11 shows the performance of different networks in cross-time taste EEG recognition after combining Grad-CAM, and the network configuration is consistent with that in section 3.3. It can be seen that the Grad-CAM method can realize the selection of important taste EEG channels after being combined with CNN-CSA and AlexNet, proving the adaptability of the Grad-CAM method in selecting taste EEG channels. Furthermore, upon comparing the outcomes of various network models in cross-time taste EEG recognition, it becomes evident that the CNN-CSA model outperforms other advanced network architectures in terms of both accuracy and F1-score, which not only shows that CNN-CSA model has better adaptability to Grad-CAM method, but also proves that it has a stronger ability to mine taste EEG features.

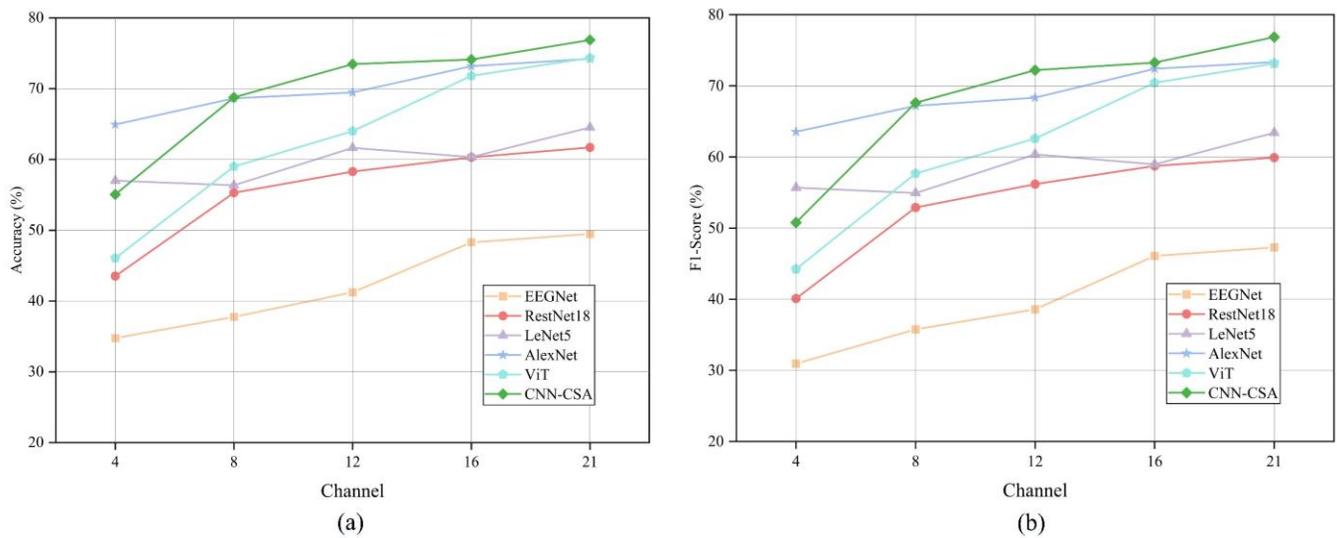

**Fig. 11.** Evaluation of various networks combined with Grad-CAM for cross-time EEG recognition: (a) accuracy, (b) F1-score.

### 3.6 Visualization

This work analyzes the spatial distribution of taste EEG features under different channel numbers using the T-SNE visualization method to visualize the effectiveness of the CAM-Attention method and CNN-CSA model. T-SNE visualization is carried out in two cases: (1) the training set and the test set are randomly divided (as shown in Fig. 12), and (2) all subjects' data from the first three days as the training set and the last day as the test set (as shown in Fig. 13). Among them, the output features of the fully connected layer in the CNN-CSA model serve as the input for T-SNE. T-SNE then transforms the features of each sample into a two-dimensional representation for visualization. In Fig. 12 and 13, the orange five-pointed star, blue square, purple triangle, and green circle represent taste EEG samples corresponding to acid, sweet, bitter, and salty categories, respectively. As can be seen from Fig. 12 and Fig. 13, when the channel number is optimized from 21 to 12, both the inter-class distance of different classes of taste EEG samples and the intra-class distance of the same classes of taste EEG samples are well maintained, and important taste EEG channels are preserved. It shows that the proposed method can improve the calculation efficiency and ensure the accuracy of taste EEG recognition. In addition, when the number of channels is reduced to 8, the feature space of taste EEG becomes worse, which shows that blindly reducing the channel number will lead to the loss of important taste EEG features. For taste EEG, 12 channels are considered for recognition efficiency and accuracy.

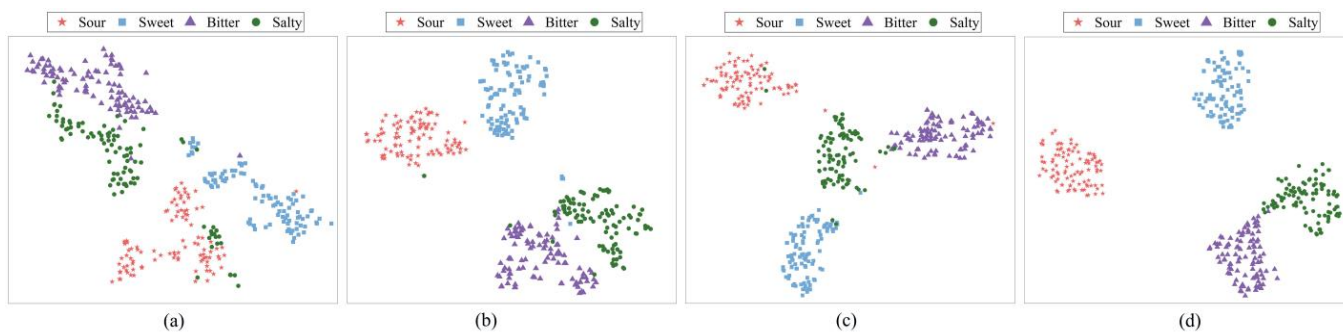

**Fig. 12.** T-SNE visualization of taste EEG feature distribution with randomly divided training set and test set: (a) 8 channels, (b) 12 channels, (c) 16 channels, (d) 21 channels.

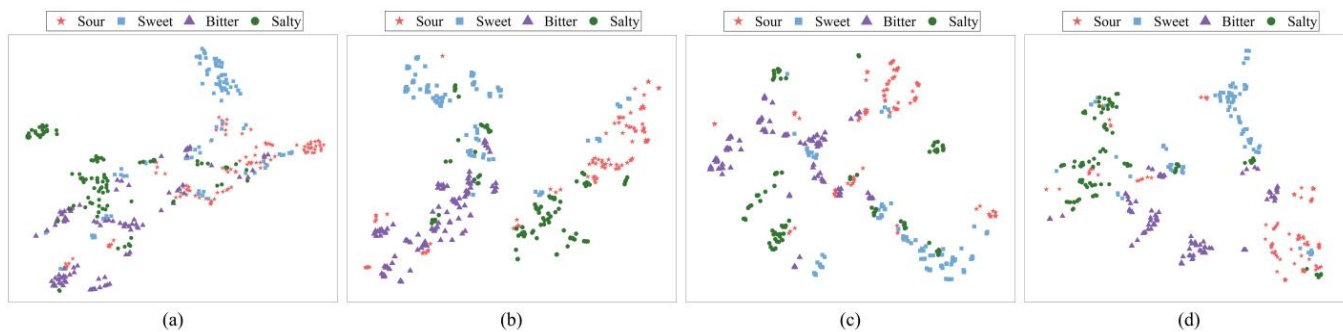

**Fig. 13.** T-SNE visualization of taste EEG feature distribution with the data of the first three days as the training set and the data of the last day as the test set for each subject: (a) 8 channels, (b) 12 channels, (c) 16 channels, (d) 21 channels.

**3.7 Potential application and limitations**

This paper proposes a channel selection method called CAM-Attention, which combines the CNN-CSA model to recognize taste EEG. Its potential application and limitations are as follows.

Potential applications. (1) Taste EEG recognition technology: This method can be applied to taste EEG recognition and provide technical support for taste research. (2) Sensory evaluation of food: This method can be used as a sensory evaluation tool of food taste and provide a reference value for the food industry. (3) Reduce the computational complexity: The CAM-Attention method can effectively reduce the computational complexity of EEG taste recognition and improve the computational efficiency.

Limitations. (1) Sample limitation: The samples of taste EEG used in this paper may be affected by experimental conditions and individual differences of subjects and may not be enough to represent the taste changes fully. (2) Generalization of the model: the effect of this method under certain taste conditions has been proven effective, but its generalization ability under other taste conditions needs further verification. Methods Generalization: The CAM-Attention method is designed for EEG recognition of taste and may not be suitable for other EEG recognition tasks. (3) Performance evaluation: Although performance indicators such as accuracy and F1 score are provided, the effect of this method in practical application is not mentioned, and more practical tests and verifications are needed.

To sum up, this method has a certain potential for application in taste EEG recognition, but some limitations still need further study and solving.

## 4. Conclusion

In this work, we propose a channel selection method of taste EEG called CAM-Attention, which combines the Grad-CAM and CNN-CSA models. The main conclusions are as follows:

(1) Firstly, we designed the taste EEG experiment and collected the taste EEG data under different taste stimuli. It provides a rich experimental database for our research and enables us to understand taste EEG's characteristics and patterns deeply.

(2) Secondly, to fully explore the key features of taste EEG, we put forward the CNN-CSA model. Through this model, we can recognize taste EEG more accurately, providing more reliable support for taste recognition.

(3) Most importantly, we combine CNN-CSA and the Grad-CAM model to form the CAM-Attention method. This method can effectively select the key channels in the taste EEG and mine the data characteristics in the selected channels, thus successfully distinguishing different tastes. This innovative method brings breakthroughs and ideas to taste EEG recognition.

To sum up, our research provides important technical support and theoretical guidance for taste EEG recognition. We believe that this method can play an important role in the study of taste EEG and provide new ideas and methods for sensory evaluation of food taste. We hope that our work can stimulate the interest of more researchers and promote the further development of this field. In the future, we will continue to work hard to improve this method further and explore a wider range of applications, making greater contributions to improve human health and quality of life.

**Declaration of Competing Interest**

The authors declare that they have no known competing financial interests or personal relationships that could have appeared to influence the work reported in this paper.

**Acknowledgements**

This work was supported by the National Natural Science Foundation of China [32471150], and the Science and Technology Development Plan of Jilin Province [YDZJ202101ZYTS135].

Xia, X., Yang, Y., Shi, Y., Zheng, W., Men, H. (2024). Decoding human taste perception by reconstructing and mining temporal-spatial features of taste-related EEGs. Applied Intelligence, 1-16. https://doi.org/10.1007/s10489-024-05374-5.

Xia, X., Guo, Y., Wang, Y., Yang, Y., Shi, Y., Men, H. (2024). Advancing cross-subject olfactory EEG recognition: A novel framework for collaborative multimodal learning between human-machine. Expert Systems with Applications 250, 123972. https://doi.org/10.1016/j.eswa.2024.123972.

Zhang, X., Yao, L., Zhang, S., Kanhere, S., Sheng, M., & Liu, Y. (2018). Internet of Things meets brain–computer interface: A unified deep learning framework for enabling human-thing cognitive interactivity. IEEE Internet of Things Journal, 6(2), 2084-2092. https://doi.org/10.1109/JIOT.2018.2877786.

Zhang, X., Li, J., Hou, K., Hu, B., Shen, J., & Pan, J. (2020, July). EEG-based depression detection using convolutional neural network with demographic attention mechanism. In 2020 42nd annual international conference of the ieee engineering in medicine & biology society (embc) (pp. 128-133). IEEE. https://doi.org/10.1109/EMBC44109.2020.9175956.

Zheng, W., Men, H., Shi, Y., Ying, Y., Liu, J., & Liu, Q. (2022). Computational model of taste pathways: a biomimetic algorithm for electronic tongue based on nerve conduction mechanism. IEEE Sensors Journal, 22(7), 6859-6870. https://doi.org/10.1109/JSEN.2022.3152057.
33